# Symplectic Tracking Using Point Magnets and a Reference Orbit Made of Circular Arcs and Straight Lines

G. Parzen

June 1993

## RHIC PROJECT

# Table of Contents

## Abstract

Symplectic tracking of beam particles using point magnets is achieved using a reference orbit made of circular arcs and straight lines that join smoothly with each other. For this choice of the reference orbit, results are given for the transfer functions, transfer matrices, and the transit times of the magnets and drift spaces. These results provide a symplectic integrator , and allow the linear orbit parameters to be computed by multiplying transfer matrices. It is shown that this integrator is a second-order integrator, and that the transfer functions can be derived from a hamiltonian.

# Symplectic Tracking Using Point Magnets and a Reference Orbit Made of Circular Arcs and Straight lines

G. Parzen

Brookhaven National Laboratory
Upton, NY 11973, USA

## 1. Introduction

In order to study long term stability, it appears desirable that the particle tracking be symplectic. One way to achieve symplectic tracking[1] is to replace the magnets by a series of point magnets and drift spaces. This approach is modified here by using a reference orbit that is made up of arcs of circles and straight lines which join smoothly with each other. This makes the symplecticity more evident, and simplifies in some way the particle tracking, as the coordinate system based on this reference orbit is not changing discontinuously between elements. It also allows the use of transfer matrices to find the linear orbit parameters. For this choice of reference orbit, the required results are obtained to track particles, which are the transfer functions, the transfer matrices and the transfer time, for the different elements present in the accelerator. It is shown that, in the absence of longitudinal magnetic fields theses results provide a symplectic, second order integrator. Existing tracking programs that use a reference orbit, made up of arcs of circles and straight lines, can be modified, using the results given here to do symplectic tracking with point magnets. The results have been used to modify the ORBIT tracking program[2]. The ORBIT program will now, by changing an indicator, either track using the usual large accelerator approximation for the transfer functions or do symplectic tracking with point magnets, and will use the same reference orbit in both cases.

## 2. Equations of Motion

The equations of motion for the transverse coordinate when no longitudinal magnetic field is present may be written as[3]

$$\frac{dx}{ds} = \frac{1 + x/\rho}{p_s} p_x \qquad (2.1a)$$

$$\frac{dp_x}{ds} = \frac{p_s}{\rho} + \frac{e}{c}(1 + x/\rho) B_y$$

$$\frac{dy}{ds} = \frac{(1 + x/\rho)}{p_s} p_y$$

$$\frac{dp_y}{ds} = -\frac{e}{c}(1 + x/\rho)B_x$$

$$p_s = \left(p^2 - p_x^2 - p_y^2\right)^{1/2}$$

x, y are the transverse coordinates in a coordinate system based on a reference orbit with radius of curvature $\rho(s)$. As the longitudinal coordinates one can use t, the particle time of arrival at s, and E the particle energy. The longitudinal coordinates obey the equations.

$$\frac{dt}{ds} = \frac{1 + x/\rho}{p_s}\frac{p}{v} \tag{2.1b}$$

$$\frac{dE}{ds} = e(1 + x/\rho)\mathcal{E}_s$$

In equation (2.1) it has been assumed that the magnetic field has no longitudinal component, $B_s = 0$, and the electric field has only a longitudinal component, $\mathcal{E}_s$. One can show that the equation for dt/ds is equivalent to, see Eq. (5.7),

$$dt = \frac{d\ell}{v} \tag{2.1c}$$

$$d\ell = \left((1 + x/\rho)^2 + (dx/ds)^2 + (dy/ds)^2\right)^{1/2}$$

where dl is the path length of the particle over ds.
The equations of motion, Eq. (2.1) may be derived from the hamiltonian

$$H = -(1 + x/\rho)\left(p^2 - p_x^2 - p_y^2\right)^{1/2} - (e/c)(1 + x/\rho)A_s \tag{2.2a}$$

where the fields are related to the vector potential $A_s$ by,

$$B_y = \frac{1}{(1 + x/\rho)}\frac{\partial}{\partial x}[(1 + x/\rho)A_s] \tag{2.2b}$$

$$B_x = -\frac{\partial}{\partial y}A_s$$

$$\mathcal{E}_s = -\frac{1}{c}\frac{\partial A_s}{\partial t}$$

It then follows that transfer functions across any element found by integrating Eq. (2.1) exactly are symplectic transfer functions. The phrase transfer functions is used here to indicate the set of functions that relates the final coordinates to the initial coordinates.

## 3. The Approximate Lattice

One procedure[1] for symplectic integration of Eq. (2.1) is to replace each magnet in the given lattice by a series of point magnets and drift spaces. The equations of motion (2.1) for the approximate lattice which has only point magnets and drift spaces can be integrated exactly when the reference orbit is made up of a series of smoothly joining arcs of circles and straight lines. This will be shown below. In addition, the result obtained by integrating the approximate lattice of point magnets and drift spaces is correct to second order in h (see section 7) where h is the distance between the point magnets, provided one chooses the strength of the point magnets as given below. Thus as one increases the number of point magnets, decreasing h, the result obtained by integrating the approximate lattice will converge to the solution of Eq. (2.1) for the given lattice. The particle motion found by integrating the approximate lattice is symplectic, as the transfer functions proposed below for the points magnets will be shown to be derivable from a hamiltonian.

## 4. Transfer Functions for Point Magnets

In the region of the lattice outside the rf cavities where the particle velocity is constant, it is convenient to use the coordinates $q_x$, $q_y$ instead of $p_x, p_y$

$$q_x = p_x/p, \quad q_y = p_y/p, \qquad (4.1)$$

$$q_s = \left(1 - q_x^2 - q_y^2\right)^{1/2} = p_s/p$$

For large accelerators $q_x \simeq dx/ds$ and $q_y \simeq dy/ds$. In the lattice region where p is constant, Eq. (2.1) can be written as

$$\frac{dx}{ds} = \frac{1 + x/\rho}{q_s} q_x \qquad (4.2)$$

$$\frac{dq_x}{ds} = \frac{q_s}{\rho} + \frac{1}{B\rho}(1 + x/\rho) B_y$$

$$\frac{dy}{ds} = \frac{1 + x/\rho}{q_s} q_y$$

$$\frac{dq_y}{ds} = -\frac{1}{B\rho}(1 + x/\rho) B_x$$

$$B\rho = pc/e \quad .$$

To construct the approximate lattice, each magnet is broken up into pieces with length h. h can be different for each piece. Each magnet piece is replaced by point magnets in one of the following ways:

1) each magnet piece is replaced by 2 point magnets at the ends of the piece separated by a drift space of length h.

2) each magnet piece is replaced by 1 point magnet in the center of the piece surrounded by drift spaces of length h/2.

Equation (4.2) suggests the following transfer functions for the point magnets. If the point magnet is located or $s = s_1$,

$$x_2 = x_1, \quad y_2 = y_1 \tag{4.3a}$$

$$q_{x2} = q_{x1} + \frac{1}{B\rho}\frac{h}{2}(1 + x_1/\rho) B_y(x_1 s_1 y_1) \quad \text{point magnets at the ends}$$

$$q_{y2} = q_{y1} - \frac{1}{B\rho}\frac{h}{2}(1 + x_1/\rho) B_x(x_1 s_1 y_1)$$

$$q_{x2} = q_{x1} + \frac{1}{B\rho}h(1 + x_1/\rho) B_y(x_1 s_1 y_1) \tag{4.3b}$$

$$q_{y2} = q_{y1} - \frac{1}{B\rho}h(1 + x_1/\rho) B_x(x_1 s_1 y_1) \quad \text{point magnet at the center}$$

It will be shown in section 7, that using the transfer functions given by Eq. (4.3), the results found using the approximate lattice are correct to order $h^2$. For the most part, results given below will be for the case where the point magnets are placed at the ends of the magnet piece.

The transfer functions in Eg. (4.3a) can be derived from the hamiltonian

$$H = -(1 + x/\rho) q_s - \frac{1}{B\rho}(1 + x/\rho)\frac{h}{2}A_s \delta(s - s_1) \tag{4.4}$$

where

$$B_y = \frac{1}{(1 + x/\rho)}\frac{\partial}{\partial x}[(1 + x/\rho) A_s]$$

$$B_x = -\frac{\partial}{\partial y}A_s$$

The transfer functions given by Eq. (4.3a) have one inconvenient feature, which is that in the case where dipoles are uniform field dipoles, the central closed orbit in a magnet piece is not in general the chord of the reference orbit between the ends of the piece. This can be corrected by introducing the factor $sin\theta/\theta, \theta = h/2\rho$, in Eq. (4.3) to give the transfer function

$$x_2 = x_1, \quad y_2 = y_1 \tag{4.5}$$

$$q_{x2} = q_{x1} + \frac{1}{B\rho}\frac{h}{2}(1+x_1/\rho)\frac{sin\theta}{\theta}B_y(x_1 s_1 y_1)$$

$$q_{y2} = q_{y1} - \frac{1}{B\rho}\frac{h}{2}(1+x_1/\rho)\frac{sin\theta}{\theta}B_x(x_1 s_1 y_1)$$

$$\theta = h/2\rho$$

Since the factor in $\sin\theta/\theta$ only changes the right side to Eq. (4.3) by terms o $(h^3)$ Eq. (4.5) is also correct up to terms of order $h^2$. Using Eq. (4.5), when the dipoles have uniform fields and the reference orbit is a circular arc, the central closed orbit is the chords of the circular reference orbit as defined by the magnet pieces.

## 5. Transfer Functions for Drift Spaces

The result for the transfer functions for drift spaces depends on the $1/\rho$ value of the reference orbit, on whether $1/\rho = 0$ or $1/\rho \neq 0$.

### 5.1 $1/\rho = 0$ drift space

Eq. (4.2) becomes

$$\frac{dx}{ds} = q_x/q_s, \quad \frac{dq_x}{ds} = 0 \tag{5.1}$$

$$\frac{dy}{ds} = q_y/q_s, \quad \frac{dq_y}{ds} = 0$$

$$q_s = \left(1 - q_x^2 - q_y^2\right)^{1/2}$$

One finds,

$$q_{x2} = q_{x1}, \quad x_2 = x_1 + q_{x1}(s_2 - s_1)/q_{s1}$$

$$q_{y2} = q_{y1}, \quad y_2 = y_1 + q_{y1}(s_2 - s_1)/q_{s1}$$

$$q_{s2} = q_{s1}$$

One may note that $(s_2 - s_1)/q_{s1}$ is just $\ell_{12}$ the path length between $s_1$ and $s_2$, as $d l = [(1 + x/\rho)/q_s] ds$.

## 5.2 $1/\rho \neq 0$ Drift Space

Eqs. (4.2) give

$$\frac{dx}{ds} = \frac{1 + x/\rho}{q_s} q_x, \quad \frac{dq_x}{ds} = \frac{q_s}{\rho} \tag{5.2}$$

$$\frac{dy}{ds} = \frac{1 + x/\rho}{q_s} q_y, \quad \frac{dq_y}{ds} = 0$$

For the $y$ motion, $q_y$ is constant and

$$q_{y2} = q_{y1} \tag{5.3}$$
$$y_2 = y_1 + q_y \ell_{12},$$

using $d\ell = [(1 + x/\rho)/q_s] ds$. A result for $\ell_{12}$ is given below.
For the $x$ motion Eq. (5.2) can be solved by using the transformation

$$q_x = a \sin \alpha \tag{5.4}$$

$$a = \left(1 - q_y^2\right)^{1/2}$$

$$q_s = \left(a^2 - q_x^2\right)^{1/2} = a \cos \alpha$$

one finds that

$$\alpha = \theta + \alpha_1, \quad \theta = (s - s_1)/\rho \tag{5.5a}$$

$$q_{x,1} = a \sin \alpha_1, \quad q_{s1} = a \cos \alpha_1$$

this gives

$$q_{x2} = q_{x1} \cos \theta + q_{s1} \sin \theta \tag{5.5b}$$

$$q_{s2} = -q_{x1} \sin \theta + q_{s1} \cos \theta$$

$$\theta = (s_2 - s_1)/\rho$$

Eq. (5.5) show that $q_x, q_s$ are rotated by the angle $\theta = (s_2 - s_1)/\rho$. The $x$ equation can be solved, using $q_x = a\sin(\theta + \alpha_1)$, $q_s = a\cos(\theta + \alpha_1)$

$$x_2 = x_1 + (1 + x_1/\rho)\ 2\rho\ \sin\theta/2\ \frac{q_{x1}\cos\theta/2 + q_{s1}\sin\theta/2}{-q_{x1}\sin\theta + q_{s1}\cos\theta} \qquad (5.6a)$$

Eq. (5.6a) can also be written as

$$x_2 = x_1 + (1 + x_1/\rho)\ 2\rho\ \sin\theta/2\ \frac{q_x(\theta/2)}{q_s(\theta)} \qquad (5.6b)$$

$$(x_2 + \rho)\,q_{s2} = (x_1 + \rho)\,q_{s1} \qquad (5.6c)$$

## 5.3 Path Length of Drift Spaces

Using

$$d\ell = \left((1 + x/\rho)^2 + (dx/ds)^2 + (dy/ds)^2\right) ds, \qquad (5.7a)$$

one finds that

$$d\ell = \frac{1 + x/\rho}{q_s} ds \qquad (5.7b)$$

For $1/\rho = 0$ drift spaces, $q_s$ is constant and

$$\ell_{12} = \frac{1}{q_{s1}}(s_2 - s_1) \qquad (5.8)$$

For $1/\rho \neq 0$, Eq. (5.7b) can be integrated using

$$\frac{\rho + x}{\rho + x_1} = \frac{q_{s1}}{q_s} = \frac{\cos\alpha_1}{\cos\theta + \alpha_1} \qquad (5.9)$$

$$q_s = a\cos(\theta + \alpha_1)$$

and gives

$$\ell_{12} = (1 + x_1/\rho) \frac{\rho \sin \theta}{q_{s2}} \tag{5.10a}$$

$$\theta = (s_2 - s_1)/\rho$$

another expression for $\ell_{12}$ is

$$\ell_{12} = \frac{2\rho \sin \theta/2}{q_s(\theta/2)} \left[1 + (x_1 + x_2)/2\rho\right] \tag{5.10b}$$

## 6. Comments on the Longitudinal Tracking

The longitudinal variables are $E$, the energy of the particle and $t$, the time of arrival at s. These are usually measured relative to the synchronous path.

To simplify things, the case discussed here is the case of the stationary bucket. In this case the particle energy in the synchronous path, $E_s$, is constant. The RF frequency, $\omega_{rf}$ is chosen so that the synchronous particle arrives at the RF cavity when the RF voltage is zero.

The particles on the synchronous path may be chosen to follow the central closed orbit of the lattice. If the lattice is made up of quadrupoles and uniform field dipoles, then for the approximate lattice, the closed orbit in the dipoles will follow the chord that join the end points of each magnet piece. Thus the path length of each magnet piece for the synchronous particle is given by

$$L_s = 2\rho \sin \theta/2 \tag{6.1}$$

$$\theta = L_m/\rho$$

$L_m$ is the magnet piece length along the reference orbit.

If the dipoles are not uniform field magnets then the central orbit of the lattice has to be computed by the tracking program which will also compute $L_s$.

Using the $L_s$ for each magnet piece, Eg. (6.1), one can compute the total path length $L$ of the approximate lattice, and then $\omega_{rf}$ is given by

$$\omega_{rf} = h_{rf} 2\pi \frac{v_s}{L} \tag{6.2}$$

where $h_{rf}$ is the chosen harmonic and $v_s$ is the velocity of the synchronous particle. Note that $\omega_{rf}$ depends on how many pieces are used for the magnets. As the number of pieces is increased, $\omega_{rf}$ will approach the correct $\omega_{rf}$ for the actual lattice. The above assumes that there is one RF cavity around the lattice.

In tracking a non-synchrous particle, one has to compute the change in $\bar{t}$ across each magnet piece. Where

$$\bar{t} = t - t_s \tag{6.3}$$

$t_s$ is the time of arrival for the synchronous particle. $\Delta \bar{t}$ across each piece is given by

$$\Delta \bar{t} = \Delta t - \Delta t_s \tag{6.4}$$

$$= \frac{L}{v} - \frac{L_s}{v_s}$$

$$\Delta \bar{t} = \frac{1}{v}(L - L_s) - L_s \left(\frac{1}{v} - \frac{1}{v_s}\right)$$

$L$ is given by Eq. (5.10) or Eq. (5.8). To avoid the cancellation that may occur, one can use the following expression for $1/\beta - 1/\beta_s$, $\beta = v/c$.

$$\frac{1}{\beta} - \frac{1}{\beta_s} = -\frac{\bar{\gamma}}{\gamma^2} \frac{(\gamma + \gamma_s)}{(\beta + \beta_s)} \frac{1}{\gamma_s \beta \beta_s} \tag{6.5}$$

$$\bar{\gamma} = (\gamma - \gamma_s)/\gamma_s, \quad \gamma = (1 - \beta^2)^{1/2}$$

Using the above result for $\Delta \bar{t}$ for each magnet piece, one can track $\bar{t}$ around the lattice. At the rf cavity the phase of the rf is given by $\omega_{rf}\bar{t}$, from which the cavity voltage and the energy gain of the particle can be computed.

## 7. A Second Order Integrator

In this section, it will be shown that the results obtained using the approximate lattice are correct to terms of order $h^2$ for each magnet piece of length $h$. The equations of motion will be written in this form

$$\frac{dx_i}{d\lambda} = f_i(x), \quad i = 1, 7 \tag{7.1a}$$

where the seventh equation is

$$\frac{ds}{d\lambda} = 1 \tag{7.1b}$$

The $fi(x)$ do not then depend on $\lambda$, which simplifies the algebra. One can then obtain the Taylor series result which is correct to order $h^2$ for the transfer functions for a magnet piece of length $s_2 = s_1 = h$.

$$x_{i2} = x_{i1} + f_{i1}h + \sum_{j=1,6} \left[\frac{\partial f_i}{\partial x_j} f_j\right]_{x_{i1}} \frac{h^2}{2} \tag{7.2}$$

$x_{i1}$ and $x_{i2}$ are the coordinate at $s = s_1$ and $s = s_2$  $f_{i1} = f_i(x_{i1})$. Eq. (7.2) can be derived from the Talyor series expansion.

$$x_{i2} = x_{i1} + \left(\frac{dx_i}{d\lambda}\right)_{x_{i1}} h + \frac{1}{2}\left(\frac{d^2 x_i}{dx^2}\right)_{x_{i1}} h^2 + ... \tag{7.3}$$

$$\left(\frac{dx_i}{d\lambda}\right)_{x_{i1}} = f_{i1}$$

$$\left(\frac{d^2 x_i}{d\lambda^2}\right)_{x_{i1}} = \left(\frac{df_i}{d\lambda}\right)_{x_{i1}} = \sum_j \left(\frac{\partial f_i}{\partial x_j}\right)_{x_{i1}} \frac{dx_j}{d\lambda} = \sum_j \left(\frac{\partial f_i}{\partial x_j} f_j\right)_{x_{i1}}$$

The transfer functions given by Eq. (7.2) are correct up to order $h^2$. Thus it is necessary to show that the approximate lattice gives the same result, up to terms of order $h^2$ as Eq. (7.2). Two approximate latices have been studied. The first where the point magnets are put as the end of each magnet piece, and the second where the point magnet is put at center of the magnet piece.

## 7.1 Point Magnets at the Ends

The equation of motion may be written as

$$\frac{dx_i}{d\lambda} = g_i + K_i \tag{7.4}$$

$$f_i = g_i + K_i$$

and $K_i \to o$ when $B_x, B_y \to o$

It may be seen from Eq. (2.1) that in the absence of longitudinal fields, $K_i = o$ when $i = 1, 3$ and $K_2$ and $K_4$ do not depend on $q_x$, $q_y$. In drift spaces, $K_i = o$, and the motion is determined by the $g_i$.

The $x_i$ begin at $x_{i1}$. After the kick due to the point magnet at $s = s_1$, the $x_i$ change to $x_{i2}$. The particle then drifts and just before the kick at $s = s_2$ the $x_i$ become $x_{i3}$. The transfer function relates $x_{i4}$ to $x_{i1}$, where $x_{i4}$ are the coordinates after the kick at $s_2$.

For the kick at $s_1$, one writes

$$x_{i2} = x_{i1} + \Delta_{i1} \tag{7.5}$$

$$\Delta_{i1} = \frac{1}{2} hK_{i1}$$

After the drift space Eq. (7.2) gives $x_{i3}$

$$x_{i3} = x_{i2} + g_{i2} h + \sum_j \left(\frac{\partial g_i}{\partial x_j} g_j\right)_{x_{i2}} \frac{h^2}{2} \tag{7.6}$$

The last term in Eq. (7.6) can be evaluated at $x_{i1}$ instead of $x_{i2}$ with an error of order $h^3$. The $g_{i2}$ can be evaluated as

$$g_{i2} = g_i\left(x_{i1} + \frac{h}{2} K_{i1}\right) \tag{7.7}$$

$$= g_{i1} + \sum_j \left(\frac{\partial g_i}{\partial x_j} K_j\right)_{x_{i1}} \frac{h}{2}$$

Thus one finds

$$x_{i3} = x_{i1} + h\left(g_{i1} + \frac{1}{2}K_{i1}\right) + \frac{h^2}{2}\left(\frac{\partial g_i}{\partial x_j} g_j + \frac{\partial g_i}{\partial x_j} K_j\right)_{x_{i1}} \tag{7.8}$$

The last kick gives

$$x_{i4} = x_{i3} + \frac{1}{2} h K_{i3} \tag{7.9}$$

$$K_{i3} = K_{i1} + \sum_j \left(\frac{\partial K_i}{\partial x_j}\right)_{x_{i1}} h\left(g_{j1} + \frac{1}{2}K_{j1}\right) + 0(h^2)$$

$$x_{i4} = x_{i1} + h(g_{i1} + K_{i1}) + \frac{h^2}{2} \sum_j \left(\frac{\partial g_i}{\partial x_j} g_j + \frac{\partial g_i}{\partial x_j} K_j + \frac{\partial K_i}{\partial x_j} g_j + \frac{\partial K_i}{\partial x_j} \frac{K_j}{2}\right) X_{i1}$$

$$x_{i4} = x_{i1} + h f_{i1} + \frac{h^2}{2} \sum_j \left(\frac{\partial f_i}{\partial x_j} f_j - \frac{\partial K_i}{\partial x_j} \frac{K_j}{2}\right)_{x_{i1}} \tag{7.10}$$

Eq. (7.10) agrees with the Taylor series result Eq. (7.2) except for the term

$$-\frac{h^2}{2} \sum_j \left(\frac{\partial K_i}{\partial x_j} \frac{K_j}{2}\right)_{x_{i1}} \tag{7.11}$$

In the absence of longitudinal fields this term is zero as $K_1 = K_3 = o$ and $\partial K_i/\partial x_j = o$ when $j = 2, 4$. The result found using point magnets at the ends of each piece agrees with the Taylor series result up to terms of order $h^2$.

In the same way, it can be shown that the approximate lattice with a point magnet at the center of each magnet piece also give a transfer function that is correct up to terms of order $h^2$.

## 8. Transfer Matrices for the Approximate Lattice

For the lattice without RF, the linear orbit parameters can be found using transfer matrices. For a magnet piece which goes from $s_1$ to $s_2$, and the coordinates go from $x_{i1}$ to $x_{i2}$, the transfer matrix is defined as

$$T_{ij} = \frac{\partial g_i}{\partial x_{j1}} \tag{8.1a}$$

where the $g_i$ are the transfer functions

$$x_{i2} = g_i(x_{i1}) \tag{8.1b}$$

The right hand side of Eq. (8.1a) is evaluated on the closed orbit for a certain $\Delta p/p$, and $i = 1, 4$ and $j = 1, 4$.

For the reference orbit used here, where the slope of the reference result does not change abruptly, the transfer matrix for one turn is just the product of the transfer matrices for each magnet piece. The linear orbit parameters, the tune, beta functions, etc. can be computed from the one turn transfer matrix.

### 8.1 Transfer Matrices for the Point Magnets

For the point magnets that replace the magnets in the approximate lattice, the transfer functions Eq. (4.5) can be used to compute the transfer matrices. One can see that $T_{ij} = 1$ *for* $i = j$, and the only other non zero $T_{ij}$ are $T_{21}$, $T_{23}$ and $T_{41}$, $T_{43}$. Thus

$$T_{ij} = \begin{bmatrix} 1 & 0 & 0 & 0 \\ T_{21} & 1 & T_{23} & 0 \\ 0 & 0 & 1 & 0 \\ T_{41} & 0 & T_{43} & 1 \end{bmatrix} \tag{8.2a}$$

$$T_{21} = \frac{\sin(\theta/2)}{\theta/2} \frac{h}{2} \frac{1}{B\rho} \left[(1+x/\rho)\frac{\partial B_y}{\partial x} + \frac{1}{\rho} B_y\right]$$

$$T_{23} = \frac{\sin(\theta/2)}{\theta/2} \frac{h}{2} \frac{1}{B\rho} (1+x/\rho)\frac{\partial B_y}{\partial y} \quad (8.2b)$$

$$T_{41} = -\frac{\sin(\theta/2)}{\theta/2} \frac{h}{2} \frac{1}{B\rho} \left[(1+x/\rho)\frac{\partial B_x}{\partial x} + \frac{1}{\rho} B_x\right]$$

$$T_{43} = -\frac{\sin(\theta/2)}{\theta/2} \frac{h}{2} \frac{1}{B\rho} (1+x/\rho) \frac{\partial B_x}{\partial y}$$

$$\theta = (s_2 - s_1)/\rho, \quad h = s_2 - s_1$$

Eqs. (8.2) are for the case where the point magnets are put at the ends of each magnet piece which goes from $s_1$ to $s_2$. For $x, y, s$ one uses the coordinates of the particle just before the point magnet. If one puts one point magnet at the center of the magnet piece, then in Eq. (8.2) one replaces $h/2$ by $h$.

## 8.2 Transfer Matrices for the Drift Spaces

For the drift spaces, the transfer functions given in Eqs. (5.3), (5.6) and (5.10) can be used to compute the transfer matrix by computing the various required derivatives of the transfer functions. The transfer functions are summarized as follows for a magnet piece that goes form $s_1$ to $s_2$.

$$y_2 = y_1 + q_{y1}\ell_{12} \quad (8.3)$$

$$q_{y2} = q_{y1}$$

$$\ell_{12} = (1 + x_1/\rho)\rho \sin\theta/q_{s2}$$

$$q_{x2} = q_{x1}\cos\theta + q_{s1}\sin\theta$$

$$q_{s2} = -q_{x1}\sin\theta + q_{s1}\cos\theta$$

$$x_2 = x_1 + (1 + x_1/\rho)2\rho\sin(\theta/2)\frac{q_x(\theta/2)}{q_{s2}}$$

$$q_x(\theta/2) = q_{x1}\cos\theta/2 + q_{s1}\sin\theta/2$$

$$q_s(\theta/2) = -q_{x1}\sin\theta/2 + q_{s1}\cos\theta/2$$

$$\theta = (s_2 - s_1)/\rho$$

One can then find the $T_{ij}$ using $T_{ij} = \partial g_i/\partial x_j$ which gives

$$T_{ij} = \begin{bmatrix} T_{11} & T_{12} & 0 & T_{14} \\ 0 & T_{22} & 0 & T_{24} \\ T_{31} & T_{32} & 1 & T_{34} \\ 0 & 0 & 0 & 1 \end{bmatrix} \qquad (8.4)$$

$$For \quad 1/\rho = 0, T_{11} = T_{22} = 1, \quad T_{31} = T_{24} = 0$$

$$T_{11} = \frac{1 + x_2/\rho}{1 + x_1/\rho} = \frac{q_{s1}}{q_{s2}} \qquad (8.5)$$

$$T_{11} = 1, 1/\rho = 0$$

$$T_{12} = (1 + x_1/\rho) 2\rho \sin\theta/2 \frac{q_s(\theta/2)}{q_{s2}q_{s1}} \left(1 + \frac{q_x(\theta/2)\, q_{x2}}{q_s(\theta/2)\, q_{s2}}\right)$$

$$T_{12} = \frac{(s_2 - s_1)}{q_{s1}} \left(1 + \left(\frac{q_{x1}}{q_{s1}}\right)^2\right), \quad 1/\rho = 0$$

$$T_{14} = (1 + x_1/\rho) 2\rho \sin\theta/2 \frac{1}{q_{s2}} \left(\cos\theta \frac{q_x(\theta/2)\, q_{y1}}{q_{s2}q_{s1}} - \sin\frac{\theta}{2} \frac{q_{y1}}{q_{s1}}\right)$$

$$T_{14} = \frac{(s_2 - s_1)}{q_{s1}} \frac{q_{x1} q_{y1}}{q_{s1}^2}, \quad \frac{1}{\rho} = 0$$

$$T_{22} = q_{s2}/q_{s1}$$

$$T_{22} = 1, 1/\rho = 0$$

$$T_{24} = -(q_{y1}/q_{s1}) \sin\theta$$

$$T_{24} = 0, \quad 1/\rho = 0$$

$$T_{31} = (q_{y1}/q_{s2}) \sin\theta$$

$$T_{31} = 0, \quad 1/\rho = 0$$

$$T_{32} = \frac{q_{y1}}{q_{s2}} \ell_{12} \left(\frac{q_{x1}}{q_{s1}} \cos\theta + \sin\theta\right)$$

$$T_{32} = \ell_{12} \frac{q_{x1} q_{y1}}{q_{s1}^2}, \quad 1/\rho = 0$$

$$T_{34} = \ell_{12} \left( 1 + \frac{q_{y1}^2}{q_{s2} q_{s1}} \cos\theta \right)$$

$$T_{34} = \ell_{12} \left( 1 + q_{y1}^2 / q_{s1}^2 \right), \quad 1/\rho = 0$$

### 8.3 Large Accelerator Approximation

The transfer matrices are not used in tracking particles for long times. They are primarily used to find the tune and other linear orbit parameter. Approximations in computing the $T_{ij}$ will produce small errors in the linear orbit parameters which may be acceptable.

One interesting limit is the large accelerator approximation which is usually valid for large accelerators. This assumes that

$$x/\rho \ll 1 \tag{8.6}$$

$$q_x^2 \ll 1, \quad q_y^2 \ll 1$$

$$\theta q_x \ll 1 \quad \theta q_y \ll 1$$

$\theta$ is the bending angle of each magnet piece.

In this limit one finds that for drift spaces

$$Tij = \begin{bmatrix} 1 & T_{12} & 0 & 0 \\ 0 & 1 & 0 & 0 \\ 0 & 0 & 1 & \ell_{12} \\ 0 & 0 & 0 & 1 \end{bmatrix}$$

$$T_{12} = (1 + x_1/\rho) 2\rho \sin\theta/2, \quad \theta = (s_2 - s_1)/\rho \tag{8.7}$$